\documentclass[english,aps,pra,superscriptaddress,longbibliography,reprint,twocolumn]{revtex4-2}
\usepackage{amssymb}
\usepackage{amsmath}
\usepackage{graphicx}
\usepackage{natbib}
\usepackage{epstopdf}
\usepackage{color}
\usepackage{braket}
\usepackage{physics}
\usepackage{mathrsfs}
\usepackage{upgreek}
\usepackage{todonotes}
\usepackage[colorlinks=true, allcolors=blue]{hyperref}
\usepackage{soul}
\begin{document}
\title{Coherent control of a triangular exchange-only spin qubit}
%\title{Coherent control of an exchange-only spin qubit in a triangular geometry}

\author{Edwin Acuna}
\affiliation{HRL Laboratories, LLC, 3011 Malibu Canyon Road, Malibu, California 90265, USA}
\author{Joseph D. Broz}
\affiliation{HRL Laboratories, LLC, 3011 Malibu Canyon Road, Malibu, California 90265, USA}
\author{Kaushal Shyamsundar}
\affiliation{HRL Laboratories, LLC, 3011 Malibu Canyon Road, Malibu, California 90265, USA}
\author{Antonio B. Mei}
\altaffiliation{Present Address: Microsoft, One Microsoft Way, Redmond, Washington 98052, USA}
\author{Colin P. Feeney}
\affiliation{HRL Laboratories, LLC, 3011 Malibu Canyon Road, Malibu, California 90265, USA}
\author{Valerie Smetanka}
\altaffiliation{Present Address: SCAD Savannah, 342 Bull Street, Savannah, Georgia 31401, USA}
\author{Tiffany Davis}
\affiliation{HRL Laboratories, LLC, 3011 Malibu Canyon Road, Malibu, California 90265, USA}
\author{Kangmu Lee}
\affiliation{HRL Laboratories, LLC, 3011 Malibu Canyon Road, Malibu, California 90265, USA}
\author{Maxwell D. Choi}
\affiliation{HRL Laboratories, LLC, 3011 Malibu Canyon Road, Malibu, California 90265, USA}
\author{Brydon Boyd}
\affiliation{HRL Laboratories, LLC, 3011 Malibu Canyon Road, Malibu, California 90265, USA}
\author{June Suh}
\affiliation{HRL Laboratories, LLC, 3011 Malibu Canyon Road, Malibu, California 90265, USA}
\author{Wonill D. Ha}
\affiliation{HRL Laboratories, LLC, 3011 Malibu Canyon Road, Malibu, California 90265, USA}
\author{Cameron Jennings}
\affiliation{HRL Laboratories, LLC, 3011 Malibu Canyon Road, Malibu, California 90265, USA}
\author{Andrew S. Pan}
\affiliation{HRL Laboratories, LLC, 3011 Malibu Canyon Road, Malibu, California 90265, USA}
\author{Daniel S. Sanchez}
\affiliation{HRL Laboratories, LLC, 3011 Malibu Canyon Road, Malibu, California 90265, USA}
\author{Matthew D. Reed}
\affiliation{HRL Laboratories, LLC, 3011 Malibu Canyon Road, Malibu, California 90265, USA}
\author{Jason R. Petta}
\affiliation{HRL Laboratories, LLC, 3011 Malibu Canyon Road, Malibu, California 90265, USA}
\affiliation{Department of Physics and Astronomy, University of California -- Los Angeles, Los Angeles, California 90095, USA}
\affiliation{Center for Quantum Science and Engineering, University of California—Los Angeles, Los Angeles, California
90095, USA}

\begin{abstract}
We demonstrate coherent control of a three-electron exchange-only spin qubit with the quantum dots arranged in a close-packed triangular geometry. The device is tuned to confine one electron in each quantum dot, as evidenced by pairwise charge stability diagrams. Time-domain control of the exchange coupling is demonstrated and qubit performance is characterized using blind randomized benchmarking, with an average single-qubit gate fidelity $F$ = 99.84\%. The compact triangular device geometry can be readily scaled to larger two-dimensional quantum dot arrays with high connectivity. 
\end{abstract}

\maketitle

Gate-defined quantum dots allow the isolation of single electrons in a voltage-tunable confinement potential \cite{Kouwenhovenreview}. A high degree of charge control can be achieved in quantum dot systems, including nearest-neighbor tunnel couplings as well as on-site energies \cite{RevModPhys.75.1,BarthSim}. By varying the number of quantum dots and their electronic occupancy, a rich spectrum of spin qubits can be explored, including single-electron ``Loss-DiVincenzo" qubits \cite{PhysRevA.57.120}, two-electron singlet-triplet  qubits \cite{Petta05,PhysRevLett.89.147902}, and three-electron exchange-only (EO) qubits \cite{DiVinceEO}. High-fidelity single- and two-qubit gates have been demonstrated with all three types of spin qubits \cite{BurkardRMP}.

Most research efforts to date have focused on the implementation of linear quantum dot arrays \cite{PhysRevApplied.6.054013,Takeda2022,Philips2022}, where gate electrodes are fanned out on the surface of the semiconductor. In order to scale spin qubits to larger system sizes and increase qubit connectivity, it is crucial to extend device fabrication to large two-dimensional (2D) quantum dot arrays. Recent progress has been made in the fabrication of small 2D arrays using single- and multi-layer gate stacks \cite{PRXQuantum.2.030331,Unseld2D,Wang2024operating}, as well as cross-bar geometries, where dots in an array share common gate electrodes \cite{2209.06609}. However, there are limits to how far these devices can be scaled up without the adoption of a fabrication process allowing dense connectivity to the interior of a large array of closely packed gate electrodes. 

With an eye towards the development of larger and densely interconnected 2D quantum dot arrays, we evaluate the performance of a close-packed triangular quantum dot array fabricated on a Si/SiGe heterostructure using the single-layer etch-defined gate electrode (SLEDGE) process \cite{doi:10.1021/acs.nanolett.1c03026}. We demonstrate control of the electron occupancy in each quantum dot down to a single electron and utilize the EO qubit encoding to evaluate the coherent performance of this device architecture. With blind randomized benchmarking \cite{Andrews2019}, we extract an average single-qubit gate fidelity of $F$ = 99.84\%, which is on par with EO devices fabricated in a linear geometry \cite{HRL_2QEO}. Our work will enable the investigation of larger and tightly packed 2D arrays of spins coupled by the exchange interaction, especially when combined with the multi-layer back-end-of-line (BEOL) process made possible by the SLEDGE platform \cite{SieuHa}.

\begin{figure}[t!]
	\includegraphics[width=0.9\columnwidth]{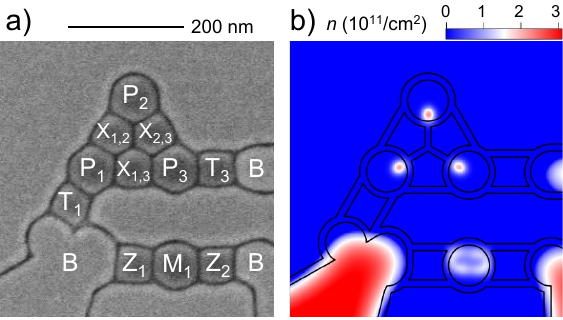}
	\caption{(a)~Scanning electron microscope image of the gate metal layer defining the close-packed triangular quantum dot array. 
Electrons are accumulated under plunger gates P$_{1}$,  P$_{2}$, and  P$_{3}$. Interdot tunnel couplings and exchange couplings are tuned with gates X$_{i,j}$, where $i$ and $j$ denote the two dots adjacent to each X gate. Gates T$_{1}$ and T$_{3}$ set the tunnel coupling to electronic reservoirs accumulated beneath the bath gates B. The quantum dot formed beneath gate M$_1$ is used as a charge sensor. (b)~Simulated charge density $n$ in the Si quantum well with the device tuned with one electron in each dot and $\sim$4 electrons in the sensor dot.}
	\label{fig_1}
\end{figure}

\begin{figure*}[t]
	\includegraphics[width=1.8\columnwidth]{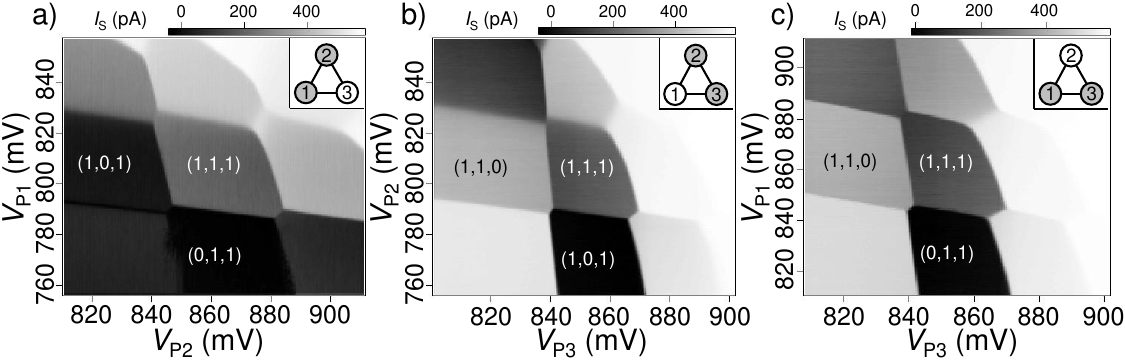}
	\caption{Pairwise charge stability diagrams. The charge sensor current $I_{\rm S}$ is plotted as a function of gate voltages $V_{P_{\rm i}}$ and $V_{P_{\rm j}}$ for a) dots 1 and 2, b) dots 2 and 3, and c) dots 1 and 3. The inset in each panel indicates the dots of the swept gates in gray and the uninvolved dot in white. Charge occupancies are denoted by ($N_1$, $N_2$, $N_3$) and the exchange-only spin qubit is operated in the (1, 1, 1) charge state. The regularity of the charging transitions is indicative of low disorder in the QW and the curvature in the charge transitions at higher electron occupancies is due to substantial interdot tunnel coupling.}
	\label{fig_2}
\end{figure*} 

The device investigated in this work is fabricated on a Si/SiGe heterostructure grown using chemical vapor deposition \cite{Deelman}. Electrons are confined in a $\sim$3~nm thick isotopically enriched (800 ppm $^{29}$Si) Si quantum well (QW) to promote a large valley splitting \cite{ChenDAPs}. The QW is buried by a 60 nm thick SiGe upper spacer layer. 2.4 nm of Al$_2$O$_3$ and 4.8 nm of HfO$_2$ serve as gate dielectrics. TiN gate electrodes are subtractively  patterned using the SLEDGE process \cite{doi:10.1021/acs.nanolett.1c03026}. These gates are then contacted vertically, from the top, using vias. These devices use a single BEOL metal layer for signal routing to the gate electrodes.

Figure 1(a) shows a scanning electron microscope image of a device nominally identical to the device under investigation. Gates P$_1$, P$_2$, and P$_3$ form a triangular quantum dot array. The exchange coupling between electrons confined in these dots is adjusted using the interdot barrier gates X$_{1,2}$, X$_{1,3}$, and X$_{2,3}$. Electrons are loaded into the array from Fermi reservoirs accumulated beneath the “bath” gates B. Gates T$_{1}$ and T$_{3}$ control the tunnel coupling to the Fermi reservoirs. The charge occupancy of the triangular quantum dot array is probed by measuring the current $I_{\rm S}$ through a charge-sensing quantum dot formed beneath gate M$_1$. The sensor dot is subject to a square-wave excitation on its source with amplitude $\sim$50 $\mu$V and the measurement integration time is $\sim$25 $\mu$s when reading out the qubit. For high-sensitivity charge detection, $I_{\rm S}$ is amplified at cryogenic temperatures and digitally demodulated at room temperature \cite{PRXQuantum.3.010352}.
 
In Fig.~1(b), we plot the simulated electron density $n$ in the $xy$-plane of the QW, which is calculated using a full Schrodinger-Poisson device simulator with a single-band effective mass model Hamiltonian, neglecting spin-orbit coupling and valley mixing effects \cite{10.1063/5.0089350}. The computational grid has a 4 nm spacing in the $xy$-plane and a 0.25 nm spacing in the $z$-direction. The two-dimensional electron gas (2DEG) density under the bath gates is calculated using a semi-classical Thomas-Fermi method, while the electron density under each dot is calculated with an explicit solution of the Schrodinger equation for the potential induced by the gates and 2DEG, where the plunger, exchange, and tunnel gate voltages are tuned to realize an equilibrium one-electron ground state in each dot with a tunnel coupling of about 10~MHz between each pair of dots. These calculations show the electrons are displaced from under the center of each plunger gate towards the center of the device, owing to the low voltages applied to the surrounding field gates and the weaker confinement provided by the exchange gates. Nonetheless, the electrons still live under their respective plungers and can be separately electrostatically controlled in simulation, giving confidence to the device design.

Spin qubits are typically operated in the few-electron regime and we now demonstrate control over the charge occupancy with finite tunnel couplings between all quantum dots. Figure~2 shows pairwise charge stability diagrams acquired by measuring $I_{\rm S}$ as a function of two plunger gate voltages, e.g.~$V_{\rm P_{1}}$ and $V_{\rm P_{2}}$ [Fig.~2(a)]. Linear voltage compensation is applied to the M$_1$ gate during these scans to maintain optimal charge sensitivity, while all other gate voltages are held fixed. The charge occupation of the triple quantum dot is denoted ($N_1$,$N_2$,$N_3$) and we achieve single electron occupancy in the center of each pairwise charge stability diagram [Figs.~2(a--c)]. The well-defined regions of charge indicate a high degree of control over the electron occupancy at the few-electron level. Moreover, the curvature of the interdot charge transitions at higher electron numbers is a consequence of significant interdot tunnel coupling \cite{RevModPhys.75.1}. Interdot tunnel couplings exceeding 10 GHz are extracted using standard measurements of the charge occupation as a function of energy level detuning \cite{PhysRevLett.92.226801,PhysRevLett.93.186802}.

\begin{figure*}[t]
	\includegraphics[width=1.8\columnwidth]{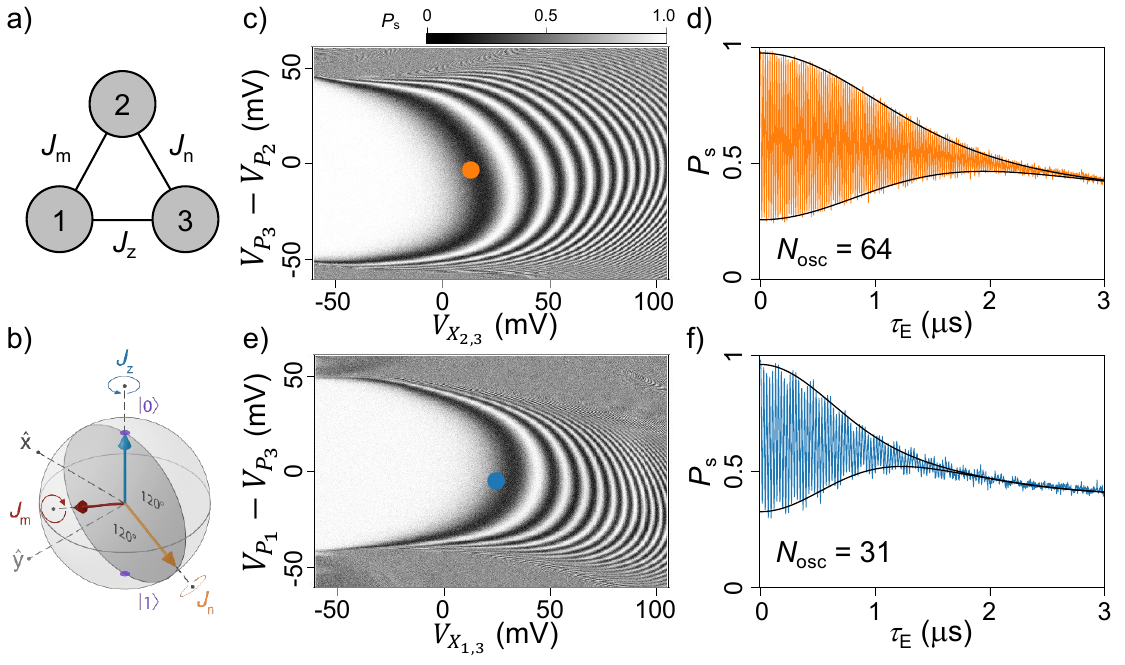}
	\caption{Coherent control at the symmetric operating point. (a)~Triangular EO spin qubit with dots 1, 2, and 3 denoted by the numbered circles. Exchange couplings $J_{\rm m}$, $J_{\rm n}$, and $J_{\rm z}$ are dynamically tuned with gates X$_{1,2}$, X$_{2,3}$, and X$_{1,3}$, respectively. (b)~Bloch sphere representation of the EO spin qubit and the rotation axes $J_{\rm m}$, $J_{\rm n}$, and $J_{\rm z}$. (c, e)~Symmetric operating point fingerprint plots demonstrating coherent rotations about the $J_{\rm n}$ axis in (c) and the $J_{\rm z}$ axis in (e). The exchange pulse duration is $\tau_{\rm E}$ = $10$ ns in both plots. (d, f)~Singlet state probability $P_{\rm S}$ measured as a function of $\tau_{\rm E}$ with operating points denoted by the colored circles in (c, e). Due to the non-orthogonal control axes of an EO qubit, a full-contrast exchange oscillation occurs between $P_{\rm S}$~=~1 and 0.25. The fitted decay envelopes (black solid lines) yield $N_{\rm osc}$ = 64 for (d) and $N_{\rm osc}$ = 31 for (f). }
	\label{fig_3}
\end{figure*}

Having demonstrated single-electron occupancy, we next tune up the device for coherent control of an EO spin qubit where the qubit is encoded in the decoherence-free subsystem (DFS) of three spins \cite{PhysRevLett.85.1758,DiVinceEO}. With the triangular configuration of the quantum dots shown in Fig.~3(a), it is possible to control the exchange interaction between each pair of spins defining the qubit. The resulting exchange Hamiltonian is
\begin{equation}
H = J_{\rm m}{\bf S}_1 \cdot  {\bf S}_2 + J_{\rm n} {\bf S}_2 \cdot {\bf S}_3 + J_{\rm z} {\bf S}_1 \cdot {\bf S}_3,
\end{equation}
where ${\bf S}_i$ is the spin of the electron isolated in dot $i$. The exchange couplings $J_{\rm m}$, $J_{\rm n}$, and $J_{\rm z}$ are controlled by the gate voltages $V_{\rm X_{1,2}}$, $V_{\rm X_{2,3}}$, and $V_{\rm X_{1,3}}$, respectively. In the Bloch sphere representation of this EO qubit, these exchange interactions can be visualized as non-orthogonal control axes separated from each other by 120$^{\circ}$ in the $xz$-plane, as illustrated in Fig.~3(b). 

In past work, EO spin qubit demonstrations were performed in linear triple quantum dot arrays where exchange coupling between the outermost spins is not possible and the Hamiltonian lacks the third exchange term \cite{Andrews2019, HRL_2QEO}. To initialize an EO qubit, a pair of electrons is prepared in a singlet state while the third electron (also referred to as the gauge electron) is left in an unpolarized spin state. Our device has three exchange couplings, but EO qubits only require control of two exchange couplings to implement single-qubit gates. We therefore have several choices for how to initialize the qubit.  
For the following demonstrations, the highest EO qubit performance was achieved by initializing and reading out using the P$_1$-P$_3$ dot pair and generating entanglement with the gauge spin via the X$_{2,3}$ gate. In the future, we plan to explore the advantages of operating the EO qubit with time-domain control of all three exchange couplings. As in past work, we follow the convention of denoting the exchange between the singlet spins as the $J_{\rm z}$ axis of rotation in the Bloch sphere representation. The exchange axis between the gauge spin ${\bf S}_2$ and spin ${\bf S}_3$ (${\bf S}_1$) is denoted by $J_{\rm n}$ ($J_{\rm m}$).

We rely on Pauli spin blockade to readout and initialize the EO qubit \cite{PhysRevApplied.12.014026,PRXQuantum.3.010352}. In this device, state preparation and readout performance improves at higher electron numbers in dot 1, and readout performance is best at the (3,1,1)-(4,1,0) charge transition \cite{Philips2022}. Singlet initialization is accomplished by thermalizing with the bath at the (4,1,0)-(5,1,0) charge transition and then separating the two electrons over the P$_1$-P$_3$ double dot \cite{PRXQuantum.3.010352}. As in prior demonstrations of EO qubits, the encoded $|0 \rangle$ state is defined as the singlet state residing in a pair of dots. 

Qubit rotations are performed via pairwise modulation of exchange and utilize the symmetric mode of operation \cite{PhysRevLett.116.110402,PhysRevLett.116.116801}. We characterize the $J_{\rm n}$ exchange landscape as shown in Fig.~3(c) by measuring the singlet probability $P_{\rm S}$ after pulsing exchange on with baseband pulses to $V_{\rm X_{2,3}}$ and evolving the state for a time $\tau_{\rm E}$ = 10 ns at various double-dot voltage detunings $V_{\rm P_3}$ – $V_{\rm P_2}$. The resulting ``fingerprint plots" are used to locate operating points minimally sensitive to gate voltage fluctuations. Figure 3(e) shows the fingerprint plot for the $J_{\rm z}$ control axis, which requires the calibration of a spin swap between P$_3$ and P$_2$. We follow exchange calibration protocols described in prior work to calibrate the spin swap as well as the angles for the EO qubit Clifford gate set \cite{Andrews2019}.

For $J_{\rm n}$, we characterize the impact of charge noise on the exchange gate by pulsing to a bias configuration with $J_{\rm n}$~$\sim$~50~MHz and measuring $P_{\rm S}$ as a function of $\tau_{\rm E}$ [Fig.~3(d)]. We perform a similar measurement for $J_{\rm z}$, with the result shown in Fig.~3(f). A Gaussian fit to the decay envelope of these oscillations yields $N_{\rm osc}$, the number of oscillations prior to a $1/e$ decay. We obtain $N_{\rm osc}$~=~64 for $J_{\rm n}$ and $N_{\rm osc}$~=~31 for $J_{\rm z}$. These results are comparable with prior characterizations of charge noise and its impact on exchange gates in SLEDGE devices \cite{HRL_2QEO}. A second exchange frequency is discernible in Fig.~3(f), which we attribute to the population of a low-lying excited state \cite{PhysRevLett.116.110402}. 

\begin{figure}[t!]
	\includegraphics[width=0.9\columnwidth]{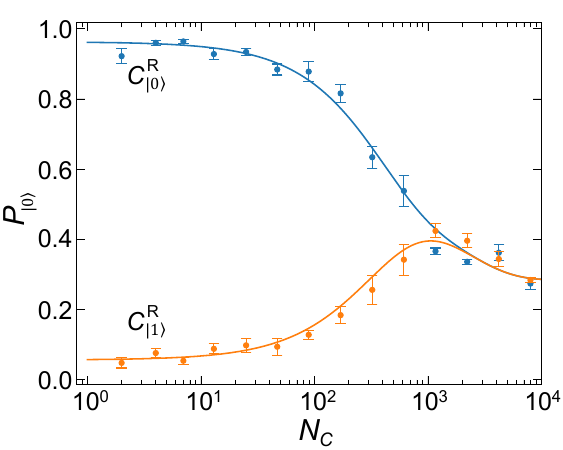}
	\caption{Characterization of the triangular EO spin qubit with BRB. The $|0 \rangle$ state return probability $P_{| 0\rangle}$ plotted as a function of the number of random Clifford gates $N_C$. The pulse sequence labeled $C^R_{|0 \rangle}$ ($C^R_{|1 \rangle}$) contains a final unitary returning the qubit to state $|0 \rangle$ ($|1 \rangle$) in the absence of errors. Fits to these data yield a single qubit gate fidelity $F$~=~99.84\%, with a total qubit error rate $\epsilon$~=~0.152~$\pm$~0.002~\% and a leakage rate $\Gamma$~=~0.024~$\pm$~0.004~\%.}
	\label{fig_4}
\end{figure}

To parameterize the performance of the triangular EO qubit we utilize blind randomized benchmarking (BRB), a quantum control verification and validation (QCVV) protocol allowing the characterization of gate errors and leakage from the encoded qubit subspace \cite{Andrews2019}. As in non-blind versions of randomized benchmarking, we begin by initializing the qubit in $|0\rangle$ and then perform a sequence of random Clifford gates. The final recovery Clifford $C_{|0 \rangle }^R$ ($C_{|1 \rangle }^R$) will return the qubit to the encoded state $|0 \rangle$ ($|1 \rangle$) in the absence of errors. We alternate and keep track of the recovery Clifford used and measure the encoded $|0 \rangle$ probability $P_{|0 \rangle}$ as a function of Clifford sequence length $N_C$. The result of such an experiment is shown in Fig.~4, where we observe two $P_{|0 \rangle}$ branches corresponding to the two recovery Clifford gates. The durations of the pulses evolving the qubit state are $\tau_{\rm E}$~=~10~ns, and are separated by an idle time of 10~ns. By fitting to the difference and sum of the two curves we extract an average single-qubit gate fidelity $F$~=~99.84\%, with a total qubit error rate $\epsilon$~=~0.152~$\pm$~0.002\% and a leakage rate $\Gamma$~=~0.024~$\pm$~0.004\%. By extrapolating to $N_C$~=~0, we also extract a state preparation and measurement fidelity of 95.3~$\pm$~0.7\% \cite{PRXQuantum.3.010352}.

In conclusion, we have characterized the operation of a triangular EO spin qubit. Charge stability diagrams show the device is capable of achieving single electron occupancy in all three dots. We demonstrate time-domain manipulation of exchange couplings with a high degree of coherence and fast $\sim$10 ns exchange gates. QCVV with BRB yields a single qubit gate fidelity $F$~=~99.84\%, which is competitive with gate performance in linear EO device architectures \cite{doi:10.1021/acs.nanolett.1c03026,HRL_2QEO}. Future work will evaluate the benefits of time-domain control of all three exchange axes, the use of Loss-DiVincenzo qubits and other qubit encodings in this geometry \cite{BurkardRMP}, and the fabrication of 2D spin qubit arrays. 

\begin{acknowledgments}
This material is based on work supported by the Army Research Office (ARO) under contract no.\ W911NF-24-1-0020 and W911NF-22-C-0002. The views and conclusions contained in this document are those of the authors and should not be interpreted as representing the official policies, either expressed or implied, of the ARO or the U.S. Government. The U.S. Government is authorized to reproduce and distribute reprints for Government purposes notwithstanding any copyright notation herein. The authors thank John Carpenter for assisting with the preparation of figures and Dr.~Charles Tahan for logistical support. We also thank Boeing Disruptive Computing \& Networks for access to test equipment.
\end{acknowledgments}

\bibliography{bib_Agro}

\end{document}